\documentclass[prl,reprint,amsmath,amssymb,aps,superscriptaddress,showpacs]{revtex4-1}

\usepackage{graphicx}
\usepackage{bm}

\begin{document}

\title{Realising Type II Weyl Points in an Optical Lattice}


\author{Kunal Shastri}
\affiliation{Division of Physics and Applied Physics, School of Physical and Mathematical Sciences, Nanyang Technological University, Singapore 637371, Singapore}

\author{Zhaoju Yang} 
\email{yang0366@e.ntu.edu.sg}
\affiliation{Division of Physics and Applied Physics, School of Physical and Mathematical Sciences, Nanyang Technological University, Singapore 637371, Singapore}

\author{Baile Zhang}
\email{blzhang@ntu.edu.sg}
\affiliation{Division of Physics and Applied Physics, School of Physical and Mathematical Sciences, Nanyang Technological University, Singapore 637371, Singapore}
\affiliation{Centre for Disruptive Photonic Technologies,
Nanyang Technological University, Singapore 637371, Singapore}


\begin{abstract}
The recent discovery of the Lorentz symmetry-violating `Type II' Weyl semimetal phase has renewed interest in the study of Weyl physics in condensed matter systems. However, tuning the exceptional properties of this novel state has remained a challenge. Optical lattices, created using standing laser beams, provide a convenient platform to tune tunnelling parameters continuously in time. In this paper, we propose a generalised two level system exhibiting type II Weyl points that can be realised using ultra-cold atoms in an optical lattice. The system is engineered using a three-dimensional lattice with complex $\pi$ phase tunnelling amplitudes. Various unique properties of the type II Weyl semimetal such as open Fermi surface, anomalous chirality and topological Fermi arcs can be probed using the proposed optical lattice scheme.
\end{abstract}


\maketitle

\par \textit{Introduction- } In-spite of its theoretical proposal as early as 1929 \cite{Weyl_1929}, Weyl Fermion, one of the intriguing particles predicted in the development of quantum field theory, has not been observed as an elementary particle. However, these massless particles have been recently detected in condensed matter experiments as low energy collective excitations, or quasi-particles, in a Weyl Semi-Metal (WSM) that supports Fermi-arc topological surface states \cite{Hassan_15_science, Hassan_15_nature, Lv_2015}. As described by quantum field theory that is intrinsically relativistic, Weyl Fermions are Lorentz covariant. Nevertheless, Lorentz symmetry can be violated in condensed matter systems, resulting in `type II' Weyl semimetal phase, that has no equivalent in quantum field theory. This newly proposed type-II WSM phase \cite{Bernevig_2015, Xu_2015}, has been experimentally observed in materials such as MoTe2 \cite{W5_2016, W12016,W22016,W32016} and LaAlGe \cite{W42016}, and has been predicted to exist in other materials such as diphosphides of Mo and W \cite{Autes_2016}.  Compared to the type I WSM phase, the type II phase exhibits many novel properties such as non-vanishing finite density of states at Fermi level \cite{Bernevig_2015} and  anomalous chiral Landau levels \cite{Shengyuan_2016, Serguei_2016, Udagawa_2016}. These properties together with the topologically protected Fermi arc surface states, apart from being interesting in fundamental physics, may be useful in the realisation of quantum computation and high efficiency circuitry. 

\par Tuning the WSM to explore its various properties has remained difficult in condensed matter systems. Alternate systems such as those based on ultra-cold atoms and classical (electromagnetic \cite{LuLing_2014} and acoustic) waves offer flexible platforms to realise and probe these WSM phases. In systems based on ultracold atoms, since the optical lattice is created using standing optical waves, parameters such as tunnelling amplitude between lattice sites can be continuously modulated in time \cite{Goldman_2016, Bloch_2008}, thereby easily tuning the properties of WSM phases. This is in contrast to condensed matter samples where tunnelling amplitude is a fixed material property. There have been a of number proposals to construct type I WSM using ultra cold atoms in optical lattices \cite{He_2016, Dubeck_2015, Ganeshan_2015, Zhang_2015, Lan_2011, Jiang_2012, Zhang_2016}. Of these, the proposals in Ref \cite{Dubeck_2015, Jiang_2012} have proved promising because they are extensions of the experimentally realised Harper-Hofstadter Hamiltonian \cite{Aidelsburger_2013, Miyake_2013} and staggered flux lattice \cite{Aidelsburger_2011}. Moreover, type I Weyl points (WPs) have been observed in photonic crystals \cite{LuLing_2015}, in photonic material based on magnetised plasma \cite{Wenlong_2016}, and predicted to exist in acoustic structures \cite{ChanCT_2015, Zhaoju_2016}.  

\begin{figure}
\includegraphics[scale=0.35]{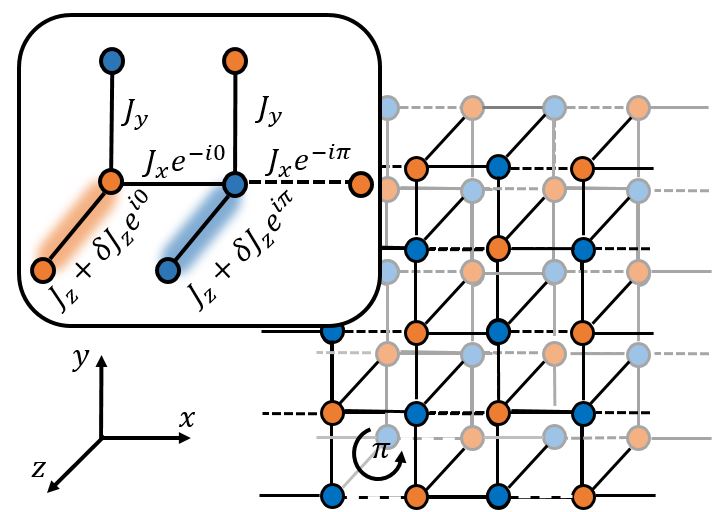}
\caption{Structure of 3D lattice proposed to realise type II Weyl points. Tunnelling amplitude along the $x$ direction alternately carries a complex phase of $\pi$ (depicted using dashed lines) and $0$ (solid). The lattice can be constructed using two dissimilar sites depicted in red and blue. Tunnelling along $z$ direction has a constant phase of $0$ ($\pi$) along red (blue) sites and is shifted from $0$ by a constant value $J_z$ Inset: A part of the lattice depicting the complex tunnelling amplitudes in the $x$, $y$ and $z$ directions.}\label{fig:Lattice}
\end{figure}

\begin{figure*}
\includegraphics[scale=0.57]{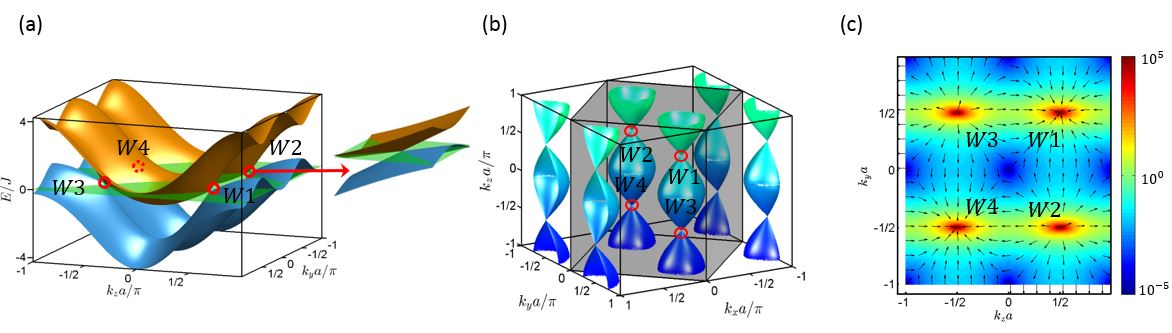}
\caption{(a) Band structure of lattice in $k_x=0$ plane. The electron (orange) and hole (blue) bands touch to form type II WPs at $(k_y,k_z)$=$(\pm\pi/2,\pm\pi/2)$. Each node, labelled $W1$-$W4$, is tilted in the $k_z$ direction. (b) Fermi surface ($E=0$) is depicted as a function of the crystal momentum $\bm{k}$. The electron-hole pockets in the first Brillouin zone (shaded in grey) touch at WPs (W1-W4). (c) Berry curvature in the $k_x = 0$ plane. Arrows represent the direction of the vector field. type II WPs $W2$, $W3$ act as sources and $W1$, $W4$ act as sinks of Berry curvature. The colour scale, indicating magnitude of the vector field, has been logarithmically scaled. In all calculations magnitude of tunnelling amplitudes is normalised to $1$ and $\delta J_z$ is assumed to be $1/2$.}\label{fig:band}
\end{figure*}

\par The type II WSM phase, on the other hand, is much less explored. Besides the recent condensed-matter realizations, this type-II WSM phase can be realised in a spin orbit coupled Bose Einstein condensate \cite{Xu_2015}, and can be constructed for electromagnetic waves in a metallic-helical-wire metamaterial system \cite{Shanhui_2016}. However, all these systems, once constructed, cannot be easily tuned. To fill this gap, we propose a model with type II WP spectra that can be realised in optical lattices using ultra-cold atoms. This model consists of a 3D square lattice with complex $\pi$ phase tunnelling amplitudes from one lattice site to another. We have chosen this model because non-zero phase tunnelling amplitudes have already been implemented in 2D ultracold atoms in optical lattices \cite{Dalibard_2011, Goldman_2016} thereby, making it easily realisable using current know-how.  

\par \textit{Model- } Consider a two dimensional lattice in the $x y$ plane, under a perpendicular magnetic flux of strength $\phi$ in each plaquette. Assuming the lattice to be deep, the system is described by the time independent Harper-Hofstadter Hamiltonian \cite{Harper_1955, Hofstadter_1976} for $\alpha=\phi/2\pi$ flux quantum per plaquette,

\begin{equation} \label{eq:1}
H=\sum_{\bm{r}}
-J_xe^{-i\Phi(\bm{r})}a^{\dagger}_{\bm{r}}a_{\bm{r}+\bm{e_x}}
-J_ya^{\dagger}_{\bm{r}}a_{\bm{r}+\bm{e_y}}+h.c
\end{equation}

Here $a^{\dagger}_{\bm{r}}$ and $a_{\bm{r}}$ are creation and annihilation operators at site $r$ and $J_x (J_y)$ are the tunnelling amplitudes in the $x (y)$ direction. $\Phi(\bm{r})$ is the position dependent complex tunnelling phase at site $\bm{r}$. The phase acquired by a particle while moving along a closed plaquette in this lattice is analogous to the Aharonov-Bohm phase gathered by the wave function of particles moving in circular trajectories under the influence of an external magnetic field, thereby simulating a synthetic magnetic field. Since the lattice is discrete, we can use integral indices $(m,n)$ to represent the position vector as $\bm{r}=a(m\bm{e_x}+n\bm{e_y})$, where $a$ is the lattice constant and $\bm{e_x}$ and $\bm{e_y}$ are unit vectors along the $x$ and $y$ direction respectively. In our model we set $\phi=\pi$ using complex hopping phases in the $x$ direction $\Phi(m,n)=\pi(m+n)$. In cold atom lattices, this system has been experimentally demonstrated in Ref \cite{Miyake_2013}. Note that this scheme preserves both the inversion and time reversal symmetry.

\par Next, we extend the lattice along the $z$ direction. To realise the type II WPs we propose to modulate the tunnelling amplitude along the $z$ direction as $J_z+\delta J_ze^{-im\pi}$ ($\delta J_z<J_z$) depending on the position along $x$ axis ($m$), thereby breaking the inversion symmetry. Alternately, this can be also viewed as a potential offset $J_z\pm\delta J_ze$ since $e^{im\pi}=(-1)^{m}$ for $m\in \mathbb{Z}$. A sketch of the proposed structure is given in Fig \ref{fig:Lattice}. The modified Hamiltonian describing this system is,

\begin{eqnarray} \label{eq:Ham}
H&=&\textstyle\sum _{m,n,l}
(- J_xe^{-i\pi(m+n)}a^{\dagger}_{m,n,l}a_{m+1,n,l}\nonumber\\*
 &-& J_ya^{\dagger}_{m,n,l}a_{m,n+1,l}\nonumber\\*
 &-& (J_z+\delta J_ze^{-i(m-1)\pi})a^{\dagger}_{m,n,l}a_{m,n,l+1} +h.c.),
\end{eqnarray}

where integers $(m,n,l)$ refer to the lattice positions along $(x,y,z)$ axis. For an infinite periodic optical lattice, we can take the Fourier transform and rewrite the Hamiltonian in quasi momentum ($\bm{k}$) space as $\sum_{\bm{k}}\bm{a}^{\dagger}_{\bm{k}}\mathcal{H}(\bm{k})\bm{a}_{\bm{k}}$. Here $\mathcal{H}(\bm{k})$, know as Bloch Hamiltonian, is given by, 

\begin{eqnarray} \label{eq:Bloch_ham}
\mathcal{H}(\bm{k})=&-& 2J_zcos(k_za)\bm{1}-2J_ycos(k_ya)\bm{\sigma_x}\nonumber\\*
&-& 2J_xsin(k_xa)\bm{\sigma_y}-2\delta J_zcos(k_za)\bm{\sigma_z}
\end{eqnarray}

In this equation $\bm{\sigma}_x, \bm{\sigma}_y, \bm{\sigma}_z$ are Pauli matrices. We can confirm that the system is symmetric under time reversal by noting $\mathcal{TH}(\bm{k})\mathcal{T}^{-1}=\mathcal{H}(\bm{k})^{*}=\mathcal{H}(-\bm{k})$, but not under inversion symmetry since, $\mathcal{IH}(\bm{k})\mathcal{I}^{-1}\ne \mathcal{H}(-\bm{k})$, where $\mathcal{T}$ and $\mathcal{I}$ are the time reversal and inversion operators. Note that in our model, type I WSM can be realised by setting $J_z=0$ and $\delta J_z=1$, as reported in \cite{Dubeck_2015}. 

\par The Hamiltonian can be briefly written as $\mathcal{H}=d_0\bm{1}+\bm{d}\cdot\bm{\sigma}$ where we have introduced the vector $\bm{d}$ with components $d_x=-2J_ycos(k_ya)$, $d_y=-2J_xsin(k_xa)$, $d_z=-2\delta J_zcos(k_za)$ and $d_0=-2J_zcos(k_za)$. The energy spectrum of the system consists of two bands given by $E=d_0\pm|{\bm{d}}|$. The two bands touch at four points $(0,\pm\pi/2a,\pm\pi/2a)$ in the first Brillouin zone (Fig \ref{fig:band}). Around these points, if the magnitude of the kinetic term $(d_0)$ in the energy spectrum exceeds the second term $(|{\bm{d}}|)$ along any direction in momentum space, the Weyl cone is tilted along this direction and the points are called `type II' WPs \cite{Bernevig_2015}. In our scheme $\delta J_z$ is always less than $J_z$ thereby tilting the Weyl spectrum in the $\pm\bm{z}$ directions and creating four type II WPs at $(0,\pm\pi/2a,,\pm\pi/2a)$ in the first Brillouin zone (Fig \ref{fig:band}a). At $\delta J_z=J_z$ there is a transition between type I and type II WSM and the particle-hole pockets are straight lines connecting the touching points in momentum space. 

\par To further confirm that the four touching points of the particle and hole band are indeed type II WPs, we plot the $E=0$ Fermi surface (FS) in Fig \ref{fig:band} (b). It can be seen that the FS is open, consisting of particle pockets (for $\pi/2a<k_z<\pi/2a$) and hole pockets (for $k_z>\pm\pi/2a$) unlike the point like FS of type I WSM \cite{Dubeck_2015}. The open FS is indicative of a finite density of states at Fermi level in contrast to the vanishing DOS for type I WPs. The nodes at the intersection of the particle-hole pockets in the first Brillouin zone (marked $W1-W4$) correspond to the location of the type II WPs in $\bm{k}$ space. 

\par Weyl points, both type I and II, always occur in pairs with opposite chirality. One of each pair is a source and the other, a sink of Berry curvature. For each band with eigenfunction $\left|n(\bm{k})\right\rangle $, the Berry curvature is defined as the curl of the Berry connection $\bm{\Omega}_n(\bm{k}))= \nabla\times\left\langle n(\bm{k}) | \nabla |  n(\bm{k})\right\rangle$. For a two level system, the Berry curvature for the hole (lower) band ($E=d_0-|{\bm{d}}|$) is given by \cite{Bernevig_2013_book},
\begin{eqnarray} \label{eq:Berry_curvature}
\Omega_{-ij}(\bm{k})= \frac{1}{2|\bm{d}|^3}\bm{d}\cdot\partial_i\bm{d}\times\partial_j\bm{d},
\end{eqnarray}

where the components of $\bm{d}$ for our system are defined in the previous section. For the scheme under consideration, we calculate Berry curvature field using Eqn. \ref{eq:Berry_curvature}. The field in the $k_x=0$ plane is plotted in Fig \ref{fig:band}(c). Points $W1-W4$ correspond to the location of the type II WPs with $W2$ and $W3$ acting as sources and $W1$ and $W4$ acting as sinks of Berry curvature. We can find the topological charge of each point by integrating the Berry curvature over a surface enclosing the point in $k$ space. The calculated topological charge for $W1$ and $W4$ is $-1$ and $W2$, $W3$ is $+1$ thereby confirming their opposite chiralities.  

\begin{figure}
\includegraphics[scale=0.6]{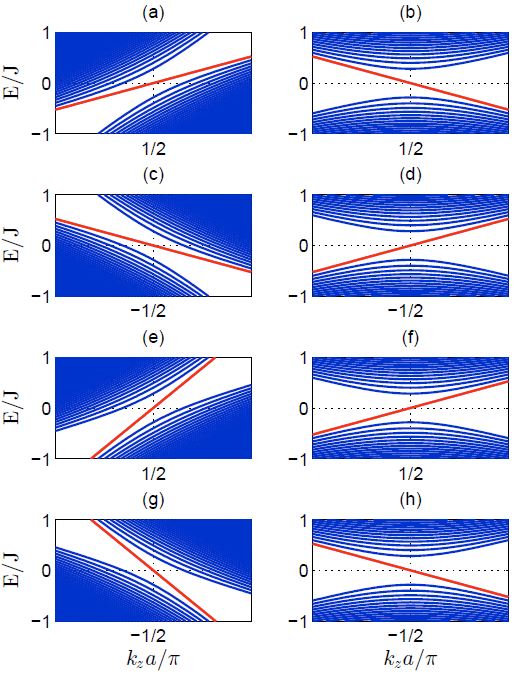}
\caption{Energy spectra in the vicinity of WPs in the presence of magnetic field ($l_B=10$) directed along $\bm{z}$ axis. First 50 Landau levels are plotted for type II WPs ($\delta J_z=0.5$ and $J_x=J_y=J_z=1$) located at $(k_x,k_y,k_z)=(0,\pi/2,\pi/2)$ in (a), $(0,\pi/2,-\pi/2)$ in (c), $(0,-\pi/2,\pi/2)$ in (e) and $(0,-\pi/2,-\pi/2)$ in (g). For comparison, spectra of corresponding type I WPs ($\delta J_z=J_x=J_y=1$ and $J_z=0$) is plotted in (b),(d),(f) and (h). The group velocities of the chiral Landau level in type I WPs (b),(f) have opposite slopes reflecting their opposite chirality. However chiral Landau levels in type II WPs (a),(e) have positive group velocities in-spite of their opposite chiralities. Same for (c,g) in contrast to (d,h)}\label{fig:Chiral_anomaly}
\end{figure}

\par \textit{Chiral Landau levels-} In the presence of a magnetic field the Landau level spectrum near a WP exhibits chiral (unpaired) zeroth level. In type I WPs the sign of the group velocity of the chiral Landau level changes with  chirality of the WP. However, in type II WPs the sign of the group velocity is independent of the chirality of the WP \cite{Serguei_2016, Udagawa_2016}. Furthermore, Landau level quantisation is possible only if the magnetic field is applied along the direction of tilt of the WPs \cite{Shengyuan_2016, Serguei_2016, Udagawa_2016}. To study these properties in the proposed structure, we first linearise the Bloch Hamiltonian (eqn. \ref{eq:Bloch_ham}) around the WPs $K=(0,\pm\pi/2,\pm\pi/2)$ to get,

\begin{eqnarray} \label{eq:Linear_ham}
\mathcal{H}(\bm{K}+\bm{q})=&\pm& 2J_zq_za\bm{1}\pm 2J_yq_ya\bm{\sigma_x}\nonumber\\*
&-&2J_xq_xa\bm{\sigma_y}\pm 2\delta J_zq_za\bm{\sigma_z}
\end{eqnarray}

\par In this condensed equation the sign of the $q_y$ and $q_z$ terms depends on the sign of $k_y$ and $k_z$ respectively, corresponding to the location of the WP ($\bm{K}$) around which the Hamiltonian is linearised. Next, we consider the model Hamiltonian in the presence of a magnetic field of strength $B$ along the $\bm{z}$ axis. The energy spectrum can be calculated by introducing the usual ladder operators $a^{\pm}=(q_x\pm iq_y)l_b/\sqrt{2}$, where magnetic length ($l_b$) is defined as $l_b=\sqrt{eB/\hbar}$. The calculated result for all four WPs with $l_B=10$ and $\delta J_z=0.5$ has been plotted in Fig \ref{fig:Chiral_anomaly} (a,c,e,g). Since $k_z$ corresponds to the tilt direction of the WP in our energy spectrum (Fig \ref{fig:band}(a)), an asymmetric chiral Landau level can be seen at each WP. Instead, if the magnetic field is induced outside the so called `magnetic' regime \cite{Serguei_2016}, say, along the $x$ and $y$ directions, no Landau levels will be observed.  Also, in Fig \ref{fig:Chiral_anomaly} (b,d,f,h) the energy spectrum for $J_z=0$ and $\delta J_z=1$, corresponding to type I WPs \cite{Dubeck_2015} has been plotted for comparison. It can be confirmed that the slope of the zeroth Landau level for WPs with opposite chirality does not change in the type II WSM in contrast to type I.

\begin{figure}
\includegraphics[scale=0.55]{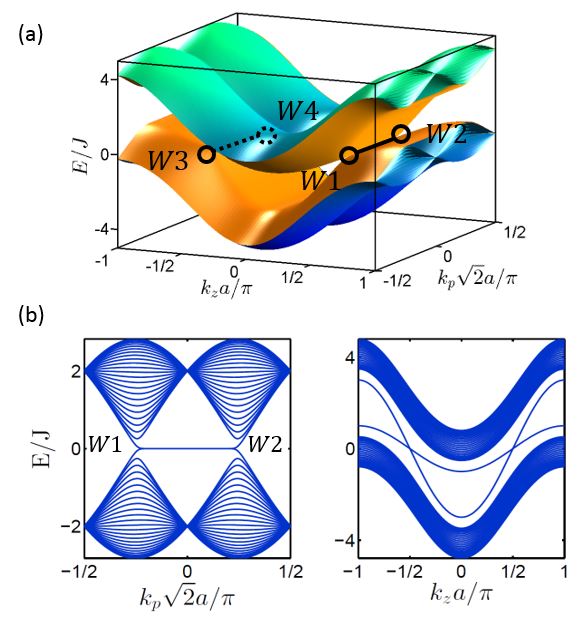}
\caption{Energy spectra for optical lattice consisting of $25$ unit cells in $\bm{x}-\bm{y}$ direction and infinite along $\bm{x}+\bm{y}$ and $\bm{z}$ direction (coordinates in Fig 1). (a) Spectra plotted as a function of $k_z$ and $k_p=(k_x+k_y)/\sqrt{2}$. Surface states have been depicted in orange. Fermi arcs connecting type II WPs $W1$ ($W3$) and  $W2$ ($W4$) are highlighted using black lines. (b) Left: spectra plotted as a function of $k_p$ for $k_z=\pi/2a$. Fermi arc connecting $W1$ and $W2$ can be clearly seen. Right: spectra plotted as a function of $k_z$ for $k_p=\pi/8a$. Two surface states corresponding to propagation along the two edges of the lattice are evident}\label{fig:Fermi_arc}
\end{figure}

\par \textit{Fermi Arcs-} One of the most exciting properties of the WSM is the existence of topologically protected surface states. In the energy spectrum these states are seen as Fermi arcs connecting WPs of opposite chirality. In order to explore the structure of Fermi arcs, in our model, we consider a lattice that is finite along the $\bm{x}-\bm{y}$ direction. We choose this direction to prevent WPs of opposite chirality from overlapping in momentum space. The energy spectrum of this finite lattice is calculated as a function of $k_z$ and $k_p=(k_x+k_y)/\sqrt{2}$ (Fig \ref{fig:Fermi_arc}). Type II WPs labelled  $W1$ ($W3$) and  $W2$ ($W4$), corresponding to opposite chiralities, are connected by Fermi arcs lying on the $k_z=\pm\pi/2a$ plane. The detailed structure of the Fermi arcs can the seen in the cross-sectional view (Fig \ref{fig:Fermi_arc}(b)) of the energy spectra. In fig \ref{fig:Fermi_arc}(b) left, $k_z$ is set to $\pi/2a$, corresponding to the location of two type II WPs (W1,W2). The Fermi arc  lying on the $k_z=\pi/2a$ plane can be seen to connect the two WPs. Two distinct surface states can be noticed when momenta is away from the WPs but lying on the Fermi arc, as in $k_p=\pi/8a$ in fig \ref{fig:Fermi_arc}(b) right.  

\par \textit{Experimental realisation-} In systems consisting of cold atoms in optical lattices many techniques can be applied \cite{Goldman_2016} to realise complex tunnelling amplitudes. Among these, the use of Raman lasers to drive resonant tunnelling between neighbouring sites has been widely used to achieve very high synthetic magnetic fields \cite{Kennedy_2015, Aidelsburger_2015, Aidelsburger_2013, Miyake_2013}. This technique, introduced in Ref \cite{Dalibard_2010, Zoller_2003, Kolovsky_2011,Lukin_2005}, is based on applying a linear energy tilt to the optical lattice and restoring the tunnelling using spatially modulated Raman lasers with detuning frequency proportional to the energy tilt. The proposed model can be realised by extending the 2D optical lattice used to demonstrate the Harper Hofstadter Hamiltonian \cite{Aidelsburger_2013, Miyake_2013}. Specifically, the 2D system in \cite{Miyake_2013} corresponds exactly to our model in the $xy$ plane. The tunnelling in $z$ direction can be implemented using an additional pair of Raman lasers in the $y z$ plane \cite{Dubeck_2015} to drive resonant tunnelling with amplitude $\delta J_z$ and phase alternating between $0$ and $\pi$ along the $x$ axis. An energy offset of $J_z$ can be created by accelerating the system along the $z$ direction. Alternate techniques based on staggered potential \cite{Aidelsburger_2011} and energy offset in nearest neighbour coupling \cite{Jotzu_2014} may also be considered to realise $z$ hopping. The type II WPs obtained using this scheme can be detected using the conventional methods of time of flight imaging and Bragg spectroscopy \cite{Bloch_2008}. Furthermore, additional pairs of lasers can be used to create a synthetic magnetic field along the $z$ direction to probe the anomalous zeroth Landau level. The proposed scheme inherits the advantages of Ref \cite{Aidelsburger_2013, Miyake_2013}, namely not requiring a mixture of spin states and applicability to a wide class of particles. 

\par Since we do not exploit the internal degrees of freedom such as spin, this concept can be extended to classical systems \cite{LuLing_2014, Zhaoju_2015} like photonics using coupled optical resonators. Non-zero phase tunnelling corresponding to large magnetic field has been experimentally implemented in these systems by changing the refractive index or optical length in a plaquette by placing the resonators asymmetrically \cite{Hafezi_2016, Hafezi_2013}. Furthermore since all non-zero phase tunnelling amplitudes in our model consist of a $\pi$ phase, corresponding to a negative real coupling $(Je^{i\pi}=-J)$, systems such as \cite{Gao_2016} can be implemented to realise the proposed structure.   

\par To summarize, in this paper we have proposed a two band model to realise type II WPs using $\pi$ phase tunnelling between adjacent lattice sites. The chosen model can be implemented using ultacold atoms in optical lattices. The proposed scheme can also be extended to other classical systems. Realization outside traditional condensed matter systems will provide tunable control over the properties of the type II WSM, thereby opening new avenues in its study.

\begin{acknowledgements}
 This work was sponsored by the NTU Start-Up Grants, Singapore Ministry of Education under Grant No. MOE2015-T2-1-070 and MOE2011-T3-1-005. We would like to thank Prof. Nicholas X. Fang and Dr. Nitin Upadhyaya for their comments and suggestions. 
\end{acknowledgements}

\par \textit{Note-} We recently came across a preprint \cite{Xu_2016_arxiv} proposing an alternative scheme to realise type II WPs using spin orbit coupling of alkali atoms.


%

\end{document}